\documentclass[showpacs,amsmath,amssymb,prl,twocolumn]{revtex4}
\usepackage{graphicx}
\usepackage{soul}
\usepackage[colorlinks=true,citecolor=blue,linkcolor=magenta]{hyperref}
\usepackage[usenames]{color}
\usepackage[cspex,bbgreekl]{mathbbol}
\usepackage{relsize}

\newcommand{\ket}[1]{\left| #1 \right>} 

\newcommand {\grsim} {\ {\raise-.5ex\hbox{$\buildrel>\over\sim$}}\ }
\newcommand {\lessim} {\ {\raise-.5ex\hbox{$\buildrel<\over\sim$}}\ } 
\newcommand {\ii} {i}

\begin{document}

\title{Experimental realization of strong effective magnetic fields in an optical lattice}

\author{M. Aidelsburger$^{1,2}$, M. Atala$^{1,2}$, S. Nascimb\`ene$^{1,2,3}$, S. Trotzky$^{1,2}$, Y.-A. Chen$^{1,2}$, and I. Bloch$^{1,2}$}

\affiliation{$^{1}$\,Fakult\"at f\"ur Physik, Ludwig-Maximilians-Universit\"at, Schellingstrasse 4, 80799 M\"unchen, Germany\\
$^{2}$\,Max-Planck-Institut f\"ur Quantenoptik, Hans-Kopfermann-Strasse 1, 85748 Garching, Germany\\
$^{3}$\,Laboratoire Kastler Brossel, CNRS, UPMC, Ecole Normale Sup\'erieure, 24 rue Lhomond, 75005 Paris, France}



\begin{abstract} 
We use Raman-assisted tunneling in an optical superlattice to generate large tunable effective magnetic fields for ultracold atoms. When hopping in the lattice, the accumulated phase shift by an atom is equivalent to the Aharonov-Bohm phase of a charged particle exposed to a staggered magnetic field of large magnitude, on the order of one flux quantum per plaquette. We study the ground state of this system and observe that the frustration induced by the magnetic field can lead to a degenerate ground state for non-interacting particles. We provide a measurement of the local phase acquired from Raman-induced tunneling, demonstrating time-reversal symmetry breaking of the underlying Hamiltonian. Furthermore, the quantum cyclotron orbit of single atoms in the lattice exposed to the magnetic field is directly revealed.
\end{abstract}

\maketitle

The application of strong magnetic fields to two-dimensional electron gases has led to the discovery of seminal quantum many-body phenomena, such as the integer and fractional quantum Hall effect \cite{tsui1982two}. Ultracold atoms constitute a unique experimental system for studying such systems in a clean and well controlled environment and for exploring new physical regimes, not attainable in typical condensed matter systems \cite{bloch2008many,fetter2009rotating}. However, charge neutrality of atoms prevents direct application of the Lorentz force with a magnetic field. An equivalent effect can be provided by the Coriolis force in a rotating atomic gas, which led to the observation of quantized vortices in a Bose-Einstein condensate \cite{madison2000vortex}. The regime of fast rotation, in which the atomic gas occupies the lowest Landau level, was achieved in Refs.~\cite{schweikhard2004rapidly} but the amplitude of the effective gauge field remained too small to enter the strongly correlated regime that requires a number of vortices on the order of the particle number \cite{cooper2001quantum,bloch2008many}. An alternative route consists in applying Raman lasers to the gas in order to realize a Berry's phase for a moving particle \cite{lin2009synthetic,dalibard2010artificial}. The effective gauge fields generated in such a setup resulted in the observation of a few vortices, but were still far from the strong-field regime.

\begin{figure}[t!]
\includegraphics[width=0.95\linewidth]{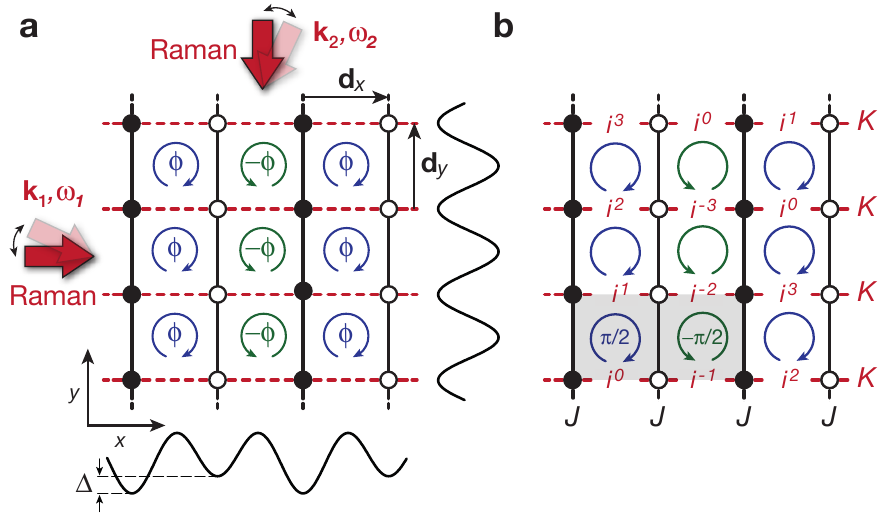}
\vspace{-0.cm} \caption{Experimental setup.
\textbf{(a)} The experiment consists of a 2D array of 1D potential tubes with spacing $|\mathbf{d}_x|=|\mathbf{d}_y|=\lambda_s/2$. 
While bare tunneling occurs along the $y$ direction with amplitude $J$, it is inhibited along $x$ owing to a staggered potential offset $\Delta$. A pair of Raman lasers with wave vectors $\mathbf{k}_{1,2}$ and frequency difference $\omega_1-\omega_2=\Delta/\hbar$, induces a resonant tunnel coupling of magnitude $K$ whose phase depends on position. This realizes an effective flux $\pm \phi$ per plaquette with alternating sign along $x$. \textbf{(b)} Spatial distribution of the phase of the Raman-induced tunnel coupling realized in the experiment. The gray shaded area highlights the magnetic unit cell.\label{Fig_Scheme} }
\end{figure}

In this Letter, we demonstrate the creation of strong effective magnetic fields for ultracold atoms in a two-dimensional optical lattice. Inspired by the proposal of Jaksch and Zoller \cite{jaksch2003creation} and subsequent work \cite{gerbier2010gauge,kolovsky2011creating,mueller2004artificial}, our technique is based on atom tunneling assisted by Raman transitions [see Fig.~\ref{Fig_Scheme}(a)]. Due to the spatial variation of the Raman coupling, the wavefunction of an atom tunneling from one lattice site to another acquires a non-trivial phase, which can be interpreted as an effective Aharonov-Bohm phase. In our setup, the magnetic flux per four-site plaquette is staggered with a zero mean, alternating between $\pi/2$ and $-\pi/2$ [see Fig.~\ref{Fig_Scheme}(b)] \cite{lim2008staggered}. We study the nature of the ground state in this optical lattice from its momentum distribution and show in particular that the frustration associated with the effective magnetic field can lead to a degenerate ground state for single particles, similar to the prediction of Ref.~\cite{moeller2010condensed}. We also study the quantum cyclotron dynamics of single atoms restricted to a four-site plaquette and obtain direct evidence for time-reversal symmetry breaking of the Hamiltonian.

\begin{figure}
\includegraphics[width=1.\linewidth]{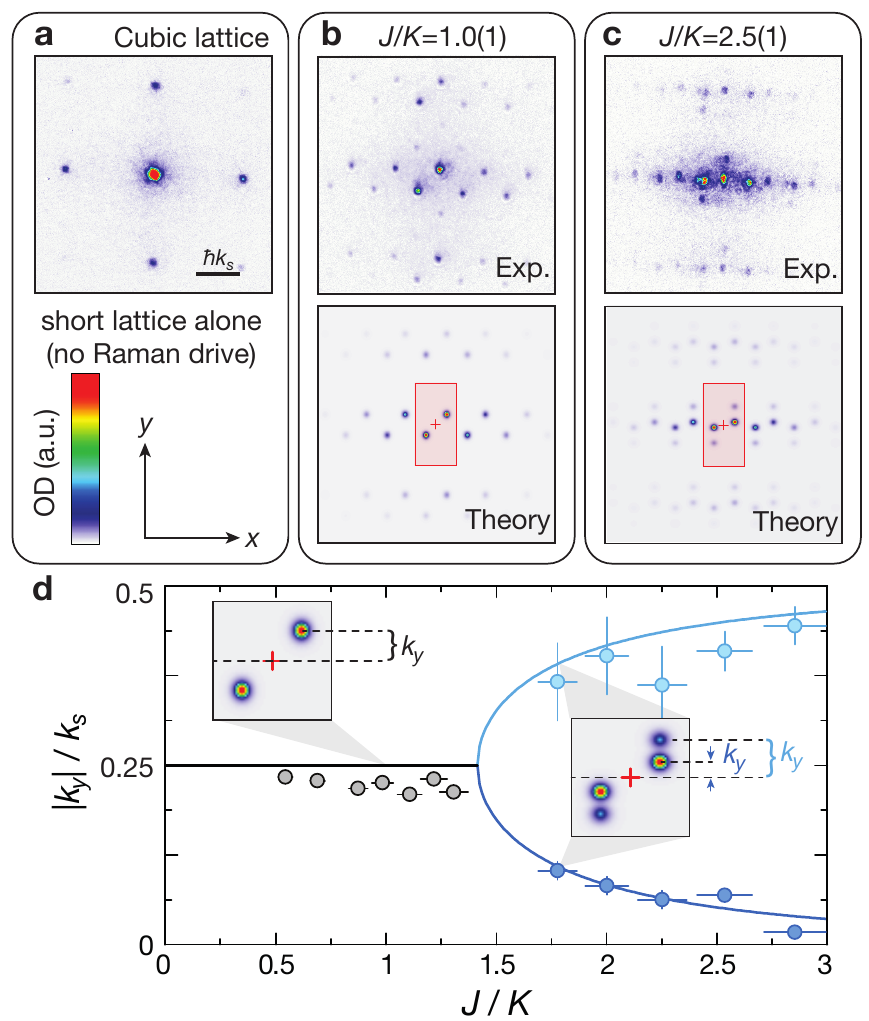}
\vspace{-0.cm} \caption{
Momentum distribution measured after a time-of-flight of $20$ ms for \textbf{(a)} simple cubic lattice, \textbf{(b)} $J/K=1.0(1)$ and \textbf{(c)} $J/K=2.5(1)$. The latter two are compared with theoretical profiles obtained by an exact diagonalization of Hamiltonian (\ref{eq_Hamiltonian}) on a $31\times 31$ lattice with harmonic confinement \cite{Supplementary}. Red squares in the theoretical profiles indicate the magnetic Brillouin zone and the crosses mark the center.
\textbf{(d)} Projection along $y$ of the momentum peaks located at $k_x=+k_s/4$, as a function of $J/K$. For $J/K<\sqrt{2}$ the peaks are located at $k_y=k_s/4$, while for $J/K>\sqrt{2}$ the peaks are split due to the emergent ground-state degeneracy (see insets). The solid lines correspond to the minima of the lowest Bloch band for a translationally invariant system \cite{Supplementary}.
\label{Fig_Diffraction_Pattern}} 
\end{figure}

Our experimental setup consists of an ultracold gas of $^{87}$Rb atoms held in a two-dimensional square lattice, forming an array of 1D Bose gases. The lattice was created by two standing waves of laser light at $\lambda_s=767$nm (`short' lattices) and a third one with twice the wavelength (`long' lattice, $\lambda_l \simeq 2\lambda_s$) to generate a staggered potential with amplitude $\Delta$ as shown in Fig.~\ref{Fig_Scheme}(a). 
 A pair of Raman lasers then induced tunneling along the staggered direction. Let us consider the Raman-assisted tunneling of an atom from a site of low energy at $\mathbf{R}=m\mathbf{d}_x+n\mathbf{d}_y$ to a site of high energy at $\mathbf{R}+\mathbf{d}_x$. Assuming $\omega_1>\omega_2$,  one obtains $K(\mathbf{R})=K\mathrm{e}^{-\ii\delta \mathbf{k}\cdot\mathbf{R}}$, where $\delta \mathbf{k}=2\pi/\lambda_K(\mathbf{e}_{\mathbf{k}_1}-\mathbf{e}_{\mathbf{k}_2})$ denotes the wave vector \cite{Supplementary}. The system is then effectively described by the non-interacting Hamiltonian
\begin{eqnarray}
\hat H=-\sum_{\mathbf{R}}\left(K\mathrm{e}^{\pm \ii \delta\mathbf{k}\cdot\mathbf{R}}\hat a^{\dagger}_{\mathbf{R}}\hat a^{\phantom{\dagger}}_{\mathbf{R}+\mathbf{d}_x}
                     +J\hat a^{\dagger}_{\mathbf{R}}\hat a^{\phantom{\dagger}}_{\mathbf{R}+\mathbf{d}_y}\right)+\mathrm{h.c.}\label{eq_Hamiltonian},
\end{eqnarray}
where the sign of the phase factor is positive for even sites of $x$ and negative otherwise. 

The phase factors in $K(\mathbf{R})$ can be interpreted as Aharonov-Bohm phases.  For the propagation of the Raman beams shown in Fig.~\ref{Fig_Scheme}(a) along $x$ and $-y$ and $\lambda_K=\lambda_l$, we obtain a phase factor of $\delta\mathbf{k}\cdot\mathbf{R}=\frac{\pi}{2}(m+n)$. Therefore the phase accumulated on a closed path around a plaquette is equal to $\phi=\pi/2$, alternating in sign along the $x$ direction. A different value of the flux $\phi$ could be achieved by choosing a different wavelength for the Raman lasers or by using a different angle between them, allowing for a fully tunable flux per plaquette in our setup.

Our experiment started by loading a Bose-Einstein condensate of about $5\times10^4$ atoms into a staggered 2D optical lattice as shown in Fig.~\ref{Fig_Scheme}(a) with $\Delta /\mathrm{h}=4.4(1)\,$kHz, resulting in an array of tubes with no coherence along $x$ \cite{Supplementary}. We then switched on the Raman lasers on resonance with strength $V_K^0=0.49(1)\,E_r$ to restore the coherence. In the limit $V_K^0\ll\Delta$, the amplitude of Raman induced tunnel coupling is  $K\simeq J_xV_K^0/(2\sqrt{2}\Delta)$, with $J_x$ being the bare tunnel coupling along $x$. For our experimental parameters, this yields a value of $K=2\pi\times 59(2)\,$Hz, in agreement with an independent measurement of $K=2\pi\times 61(3)\,$Hz \cite{Supplementary}.
After holding the atoms in this configuration for $10$\,ms, we observed a momentum distribution with restored phase coherence as shown in Fig.~\ref{Fig_Diffraction_Pattern}(b).  This can be attributed to a redistribution of entropy present in the random phases between the 1D condensates into their longitudinal modes.

\begin{figure}[t!]
\includegraphics[width=  \linewidth]{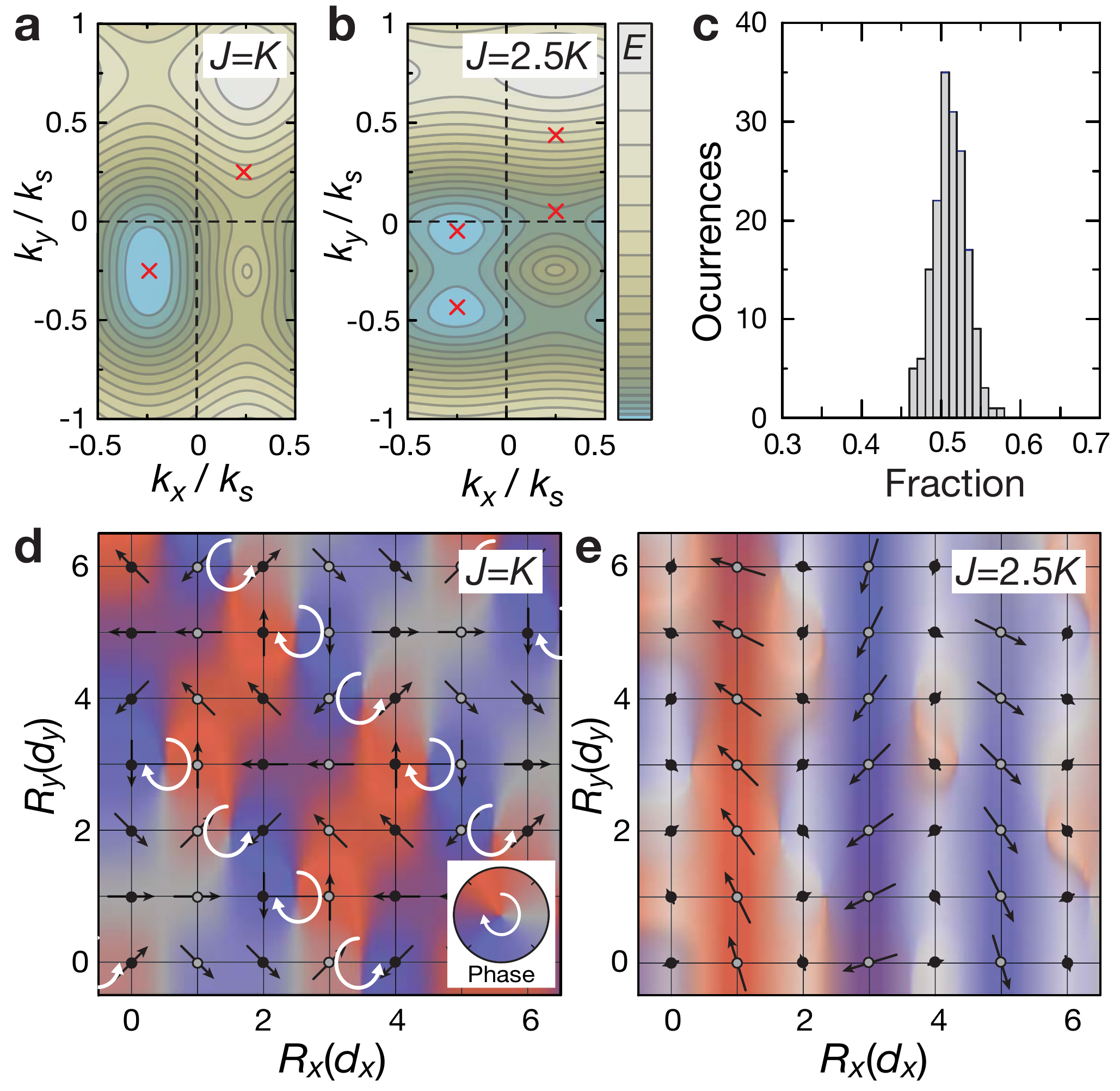}
\vspace{-0.cm} \caption{
 Dispersion relation of the lowest Bloch band, calculated in the tight-binding approximation \cite{Supplementary} for \textbf{(a)} $J/K=1$ and \textbf{(b)} $J/K=2.5$ and plotted for the first magnetic Brillouin zone. The red crosses mark the positions of the corresponding momentum peaks together with the ones shifted due to the intrinsic structure of the wavefunction.
\textbf{(c)} Histogram of the measured fraction of atoms in peaks corresponding to the lower momentum state for $J/K=2.5$. The measurement was performed over 172 identical experimental runs. 
\textbf{(d)}-\textbf{(e)} Spatial distribution of the phase and atomic density (color brightness) for the ground-state wave function \cite{Supplementary}. The vortices with different chirality in the phase distribution for $J/K=1$ \textbf{(d)} are illustrated by the rotation of the white arrows. While in this case the atomic density is uniform, it exhibits a charge density wave for $J/K=2.5$ \textbf{(e)}. For the second degenerate ground state we observe similar behavior but the density pattern is shifted by one lattice site.
\label{Fig_dispersion_flux}} 
\end{figure}

To understand the momentum distribution, we calculated the bandstructure of this lattice in the tight-binding approximation according to Ref.~\cite{moeller2010condensed}. In the presence of the gauge field, the Hamiltonian remains periodic. However, the magnetic unit cell contains two non-equivalent sites, leading to a splitting of the tight-binding band structure into two subbands \cite{Supplementary,blount1962bloch,wang2006hofstadter}. The frustration introduced by the position-dependent phase factors in $K(\mathbf{R})$ causes the phase of the atomic wavefunction to be non-uniform, leading to a condensation at non-zero momenta.  In the case $J/K = 1$, we obtain a non-degenerate ground state. However, the wavefunction itself consists of two momentum components, therefore we observe two diffraction peaks within the first magnetic Brillouin zone, one being shifted by $\Delta \mathbf{k}=(k_s/4,k_s/4)$ with respect to the minimum of the dispersion relation \cite{Supplementary}. Due to the unit cell containing more than one lattice site, the size of the  magnetic Brillouin zone is reduced compared to the one of the square lattice [see Fig.~\ref{Fig_dispersion_flux}(a)]. Therefore the ground-state momentum distribution exhibits several peaks in the short-lattice Brillouin zone, whose measured positions are in good agreement with the quasi-momenta of the Bloch states of lowest energy [see Fig.~\ref{Fig_Diffraction_Pattern}(b)]. 

When varying the ratio $J/K$ by adjusting the $y$-lattice depth, we observed the positions and number of peaks in the momentum distribution to remain unchanged for $J/K\lessim 1.4$ [see Fig.~\ref{Fig_Diffraction_Pattern}(d)]. Above this value, additional peaks appear [see Fig.~\ref{Fig_Diffraction_Pattern}(c)], which correspond to the population of two degenerate ground states. This behavior agrees with the bandstructure calculation, which shows a bifurcation at $J/K=\sqrt{2}$, above which the energy minimum is split into two degenerate ground states [see Fig.~\ref{Fig_dispersion_flux}(b)].
We find the measured and predicted peak positions of these two ground states to be in good agreement [see Fig.~\ref{Fig_Diffraction_Pattern}(d)]. 
The nature of the bifurcation is identical to the one predicted in Ref.~\cite{moeller2010condensed} where it is induced by a variation of the magnetic flux amplitude at a fixed value of $J/K=1$.  As shown in Fig.~\ref{Fig_dispersion_flux}(d), the atomic density is uniform for $J/K=1$, while in the case $J/K=2.5$ it is strongly modulated for both single-particle ground states [see Fig.~\ref{Fig_dispersion_flux}(e)]. For $J>\sqrt{2}K$ the bare coupling dominates and the phases of the atomic wavefunction tend to align along the $y$ direction, thereby frustrating the phase relation imposed by the Hamiltonian. As a consequence, the density in every second stripe along $y$ is suppressed. Contrary to the case of a triangular lattice with frustrated hopping studied in \cite{Struck21072011}, the atom fraction in each single-particle ground state does not fluctuate, as shown in Fig.~\ref{Fig_dispersion_flux}(c), and we observe an equal population in both states as predicted for interacting \cite{moeller2010condensed} or finite size systems \cite{Note}. 

\begin{figure}[t!]
\includegraphics[width=1.\linewidth]{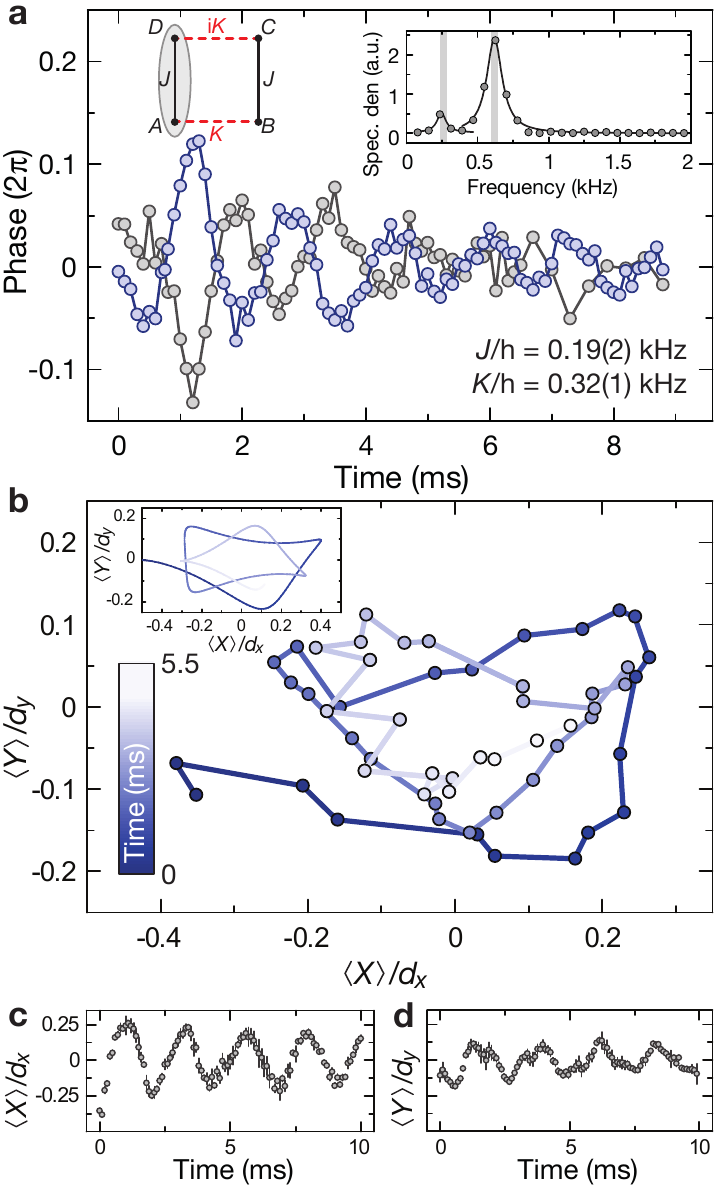}
\vspace{-0.cm} \caption{Time reversal symmetry breaking and cyclotron orbits.
\textbf{(a)} Phase evolution  of the double-slit pattern  along $y$ (integrated along $x$), as a function of time for $\hbar\omega=\Delta$ (blue) and $\hbar\omega=-\Delta$ (gray). The inset shows the Fourier transformation  for $\hbar\omega=\Delta$ depicting two frequency components at $0.24(6)\,$kHz and $0.62(13)\,$kHz, in good agreement with theory (vertical lines).
\textbf{(b)} Cyclotron orbits of the average particle position obtained from the mean atom positions $\left<X\right>/d_x$ \textbf{(c)} and $\left<Y\right>/d_y$ \textbf{(d)} for $J/\mbox{h}=0.50(2)\,$kHz, $K/\mbox{h}=0.27(1)\,$kHz and $\Delta/\mbox{h}=5.05(2)\,$kHz. Each data point is an average over three measurements. The inset in \textbf{(b)} shows the theoretical curve calculated for $\phi=0.80\times\pi/2$ and a $1/e-$damping time of $13\,$ms obtained from damped sine fits to $\left<X\right>/d_x$ and $\left<Y\right>/d_y$.
\label{Fig_Local_Phase_Measurement}} 
\end{figure}

In order to directly reveal time-reversal symmetry breaking of the Hamiltonian, we probed the local structure of the lattice with artificial gauge field at the level of a four-site square plaquette, which allows us to isolate plaquettes with equal sign of the flux. This was achieved by applying superlattice potentials along both the $x$- and $y$-directions, and in order to avoid coupling to axial modes along the potential tubes, an additional lattice along the $z$-direction ($\lambda_z=844\,$nm) was used in these measurements. The four sites of a single plaquette are denoted as $A,B,C,D$ [see inset Fig.~\ref{Fig_Local_Phase_Measurement}(a)]. The relative phases of the long and short lattices were adjusted so that the plaquettes were symmetric along $y$ and tilted along $x$, with an energy offset $\Delta/\mathrm{h}=6.0(1)\,$kHz. We first loaded single atoms in the ground state of the tilted plaquettes $\left|\psi_1\right>=(\left|A\right>+\left|D\right>)/\sqrt{2}$, and subsequently switched on the Raman lasers with $\hbar\omega=\Delta$ in order to induce resonant coupling to the $B$ and $C$ sites. In the limit $J\ll K$, the dynamics along $y$ would be suppressed and the initial state $\left|\psi_1\right>$ couples to the state $\left|\psi_2\right>=(\left|B\right>+\ii\left|C\right>)/\sqrt{2}$, where the relative phase is induced by the Raman lasers. In our case ($J/K \approx 0.5$) the evolution of the imprinted phase factors is more complex. We measured this evolution through the shape of the momentum distribution obtained after time-of-flight \cite{Supplementary}. This dynamics is a direct consequence of the complex phase factor, revealing the time-reversal symmetry breaking of the  Hamiltonian [see Fig.~\ref{Fig_Local_Phase_Measurement}(a)]. For $\omega=-\Delta/\hbar$ the role of the Raman beams is exchanged leading to a sign reversal of the phase evolution.

We also investigated this dynamics in real space in order to exhibit the influence of the gauge field on the particle flow. By generalizing the site-resolved measurement technique performed in \cite{foelling2007direct} for an array of double-wells to plaquettes, we measured the atom population per site $N_{q}$ ($q={A,B,C,D}$) \cite{Supplementary}, thus obtaining the average atom positions $\left<X\right>=(-N_A+N_B+N_C-N_D)d_x/2N$  and $\left<Y\right>=(-N_A-N_B+N_C+N_D)d_y/2N$, with $N$ being the total atom number. In the initial state $\ket{\psi_1}$ the atoms occupy the left wells $A$ and $D$ with equal weights. After switching on the Raman lasers we observe a coherent particle flow inside the plaquettes. Besides the particle current towards the right wells $B$ and $C$ [see Fig.~\ref{Fig_Local_Phase_Measurement}(c)], we observe a deviation of the mean atom position along $y$ [see Fig.~\ref{Fig_Local_Phase_Measurement}(d)]. This behavior is reminiscent of the Lorentz force acting on a charged particle in a magnetic field. As shown in Fig.~\ref{Fig_Local_Phase_Measurement}(b), the mean atom position follows an orbit that is a small-scale quantum analog of the classical cyclotron orbits for charged particles. This coherent evolution is damped due to spatial inhomogeneities in the atomic sample. Having independently calibrated the values of $J$ and $K$, we fit from the measured atom dynamics the value of the magnetic flux $\phi=0.73(5)\times\pi/2$. The difference from the value $\phi=\pi/2$ expected for a homogeneous lattice stems from the smaller distance between lattice sites inside the plaquettes when separated. For the parameters used in Fig.~\ref{Fig_Local_Phase_Measurement}(b)-(d) we calculate a distance $d_y=0.78(1)\times\lambda_s/2$ yielding $\phi=0.80(1)\times\pi/2$, which qualitatively explains the measured flux value. Residual deviations might be due to an angle mismatch between the Raman beams and the lattices beams. 

In conclusion, we have demonstrated a new type of optical lattice that realizes strong effective magnetic fields and breaks time-reversal symmetry. We have shown that the atomic sample relaxes to the minima of the magnetic bandstructure, realizing an analogue of a frustrated classical spin system. However, the spatial average of the magnetic flux is zero, hence the Bloch band is topologically trivial \cite{haldane1988model,wang2006hofstadter,gerbier2010gauge}. By using a superlattice potential with more than two non-equivalent sites \cite{gerbier2010gauge} or a linear tilt potential \cite{jaksch2003creation}, it is possible to create a lattice with a uniform and non-zero magnetic flux. This system would realize the Harper Hamiltonian \cite{harper1955single} and lead to the fractal band structure of the Hofstadter butterfly \cite{hofstadter1976energy}. In particular the lowest band would exhibit a Chern number of one and be analogous to the lowest Landau level \cite{jaksch2003creation,mueller2004artificial,dalibard2010artificial,cooper2011optical}. Our work constitutes an important step towards the study of quantum Hall effect with ultracold atomic gases and the creation of strongly-interacting liquids such as the Laughlin state \cite{sorensen2005fractional}.

We acknowledge insightful discussions with N. Cooper. This work was supported by the DFG (FOR635, FOR801), the EU (STREP, NAMEQUAM, Marie Curie Fellowship to S.N.), and DARPA (OLE program).

\section*{Appendix}

\bigskip

\renewcommand{\thefigure}{A\arabic{figure}}
 \setcounter{figure}{0}
\renewcommand{\theequation}{A.\arabic{equation}}
 \setcounter{equation}{0}
 \renewcommand{\thesection}{A.\Roman{section}}
\setcounter{section}{0}

\section{Bandstructure calculation}

We calculated the bandstructure of our lattice according to Ref.~\cite{a.moeller2010condensed}. For staggered fluxes a unit cell can be chosen as depicted in Fig.~1(b) of the main text which consists of two plaquettes with opposite sign of the flux. This cell contains two non-equivalent sites, which are denoted as even and odd. For the gauge used in our setup the following ansatz for the wavefunction can be made:
\begin{widetext}
\begin{equation}
\psi = \mathrm{e}^{\ii (m\cdot k_x d_x + n\cdot k_y d_y)} \times  
\left\{ {
   \begin{array}{ll}
    \psi_e &  \quad m\ \textrm{even}\\
    \psi_o\ \mathrm{e}^{\ii \frac{\pi}{2} (m + n)} & \quad  m \ \textrm{odd}\\
   \end{array} }\qquad, \right.
\end{equation}
\end{widetext}
with the quasi-momentum $\mathbf{k}$ in the ranges $-\pi/(2d_x) \leq k_x < \pi / (2d_x)$ and $-\pi/d_y \leq k_y < \pi/ d_y$. Substituting this ansatz into the Schr\"odinger equation using the Hamiltonian given in Eq. (1) of the main text, this reduces to a two-dimensional eigenvalue equation of the following form:
\begin{widetext}
\begin{equation}
\begin{pmatrix}
-2J\ \textrm{cos}(k_y d_y) & -K\ \mathrm{e}^{\ii k_x d_x} + \ii K \mathrm{e}^{-\ii k_x d_x} \\
-K\ \mathrm{e}^{-\ii k_x d_x} - \ii K \mathrm{e}^{\ii k_x d_x} & 2J\ \textrm{sin}(k_y d_y) \\
\end{pmatrix}
\begin{pmatrix}
\psi_e\\
\psi_o\\
\end{pmatrix}= E
\begin{pmatrix}
\psi_e\\
\psi_o\\
\end{pmatrix} \ .
\end{equation}
\end{widetext}
For the case $J/K=1$ we calculated the dispersion relation [see Fig.~3(a)] showing a single minimum located at $\mathbf{k} = (-k_s/4,-k_s/4)$. The wavefunction given in Eq. (S1), however, has two components at $\mathbf{k}$ (even sites) and $\mathbf{k}+(k_s/2,k_s/2)$ (odd sites). Therefore, in the time-of-flight measurements, we see two momentum components in the first magnetic Brillouin zone. The eigenvector in this case has equal weight on the two components, thus we see two equally populated peaks. The remaining peaks are shifted by replicas of the reciprocal lattice vectors given by $(k_s,0)$ and $(0,2k_s)$. 

For $J/K=2.5$ the ground-state contains two minima as shown in Fig.~3(b). For the same reason described above we observe two additional peaks shifted by $(k_s/2,k_s/2)$.

\section{Experimental sequence}
The experimental sequence started by loading a Bose-Einstein condensate of about $5\times10^4$ atoms with no discernible thermal fraction in 160\,ms into a 2D optical lattice of depth $V_{xs}=14\,E_r$ and $V_{ys}$ adjusted to a value, which corresponds to the desired bare tunnel coupling $J$, where $E_r=\mathrm{h}^2/(2m\lambda_s^2)$ is the lattice recoil energy. We then ramped up the long-lattice along the $x$-direction to $1.25\,E_r$ during $0.7\,$ms and subsequently decreased the short-lattice depth along $x$ to 9\,$E_r$ in $1\,$ms. The lattice potential along $x$ is then given by an alternating energy offset $\Delta=\mathrm{h}\times4.4(1)$\,kHz between neighboring sites, which was measured independently. The bare tunnel coupling $J_x=\mathrm{h}\times 190(10)\,$Hz along $x$ was much smaller than $\Delta$, resulting in an inhibition of spontaneous tunneling along $x$, therefore the initial state had no coherence along $x$.

For the case of isolated plaquettes, the experimental sequence started by loading the condensate into a 3D optical lattice of depth $V_{xl}=35\,E_r^{l}$, $V_{yl}=35\,E_r^{l}$ and $V_{z}=30\,E_r^{z}$, where $E_r^{i}=\mathrm{h}^2/(2m\lambda_i^2), i=l,z$. After that, a filtering sequence was applied to have at most one atom per lattice site \cite{a.foelling2007direct}. By ramping up the short-lattice along the $x$-direction during $10\,$ms with the desired energy offset $\Delta$, a double well potential was created with single atoms localized on one side. Finally the short-lattice along the $y$-direction was ramped up during $1\,$ms to create a plaquette potential where the atom is in the initial state $(\vert A\rangle + \vert D\rangle)/\sqrt{2}$.

\section{Raman-assisted tunneling}
In our experiments tunneling is inhibited along one direction of the lattice due to an energy offset $\Delta$. By using a pair of Raman lasers tunneling can be induced through an additional time-dependent potential $V_{K}(\mathbf{R},t)=V_K^0[\mathrm{e}^{\ii(\delta \mathbf{k}\cdot\mathbf{R}-\omega t)}+\mathrm{e}^{-\ii(\delta \mathbf{k}\cdot\mathbf{R}-\omega t)}]$, where $V_K^0$ is controlled by the Raman beam intensities, $\mathbf{R}=m \mathbf{d}_x+n \mathbf{d}_y$ is the lattice vector, $\delta \mathbf{k}=2\pi/\lambda_K(\mathbf{e}_{\mathbf{k}_1}-\mathbf{e}_{\mathbf{k}_2})$ denotes the wave vector and $\omega=\omega_1-\omega_2$ the frequency difference of the Raman lasers. If we consider an atom tunneling from a site of low energy at $\mathbf{R}$ to a site of high energy at $\mathbf{R}+\mathbf{d}_x$, and assuming $\omega_1>\omega_2$, the complex tunneling matrix element is given by the overlap integral:
\begin{eqnarray}
K(\mathbf{R})&=&V_K^0\int\mathrm{d}^2r\, w^*(\mathbf{r}-\mathbf{R})w(\mathbf{r}-\mathbf{R}-\mathbf{d}_x)\mathrm{e}^{\ii\delta \mathbf{k}\cdot\mathbf{R}}\nonumber\\
             &=&K\mathrm{e}^{\ii \delta \mathbf{k}\cdot\mathbf{R}}\label{Raman_coupling}.
\end{eqnarray} 
Here $w(\mathbf{r})$ is the Wannier function for $\mathbf{R}=\mathbf{0}$ and $K\equiv K(\mathbf{0})$ can be chosen real by convention. In the case of Raman-assisted tunneling from a high site to a low site, one obtains $K(\mathbf{R})=K\mathrm{e}^{-\ii\delta \mathbf{k}\cdot\mathbf{R}}$ and the system can be described by the effective Hamiltonian given in Eq.~(2) in the main text.

We also used Raman-assisted tunneling to characterize our system. 
To measure the value of the induced tunnel coupling $K$, we performed an independent measurement in isolated double wells. The parameters of this system $\Delta$ and $J_x$ are chosen to match the ones of the staggered potential. After loading single atoms on the lower site of the double wells we induced tunneling by the Raman beams. The atoms then undergo tunnel oscillations between the two sites of the double well whose frequency is equal to $2K$. 

The energy offset $\Delta$ can be measured starting from the same initial state by scanning the frequency difference $\omega$ of the two Raman beams. When $\hbar\omega=\Delta$ one can observe particle transfer to the second site.

\section{Data analysis for Phase measurement}

In a system of isolated symmetric double wells, where we denote the two sites as $\ket{L}$ and $\ket{R}$, the phase $\varphi$ of an arbitrary single particle state 

\begin{equation}
\frac{1}{\sqrt{2}}(\ket{L}+\mathrm{e}^{i\varphi}\ket{R})
\end{equation}

can be revealed from its momentum distribution after time-of-flight $t_{TOF}$. The observed density distribution for a sufficiently large $t_{TOF}$ is proportional to $\mathrm{cos}(k y+\varphi)$ times an envelope determined by the Wannier function, where $k=md/\hbar t_{TOF}$, $d$ is the distance between the two sites and $m$ is the atomic mass.

For the case of isolated plaquettes we first integrate the momentum distribution along $x$ and then determine the phase $\varphi$ along $y$.

\section{Site-resolved detection}
\begin{figure}[thb]
\begin{center}
\includegraphics[scale=1]{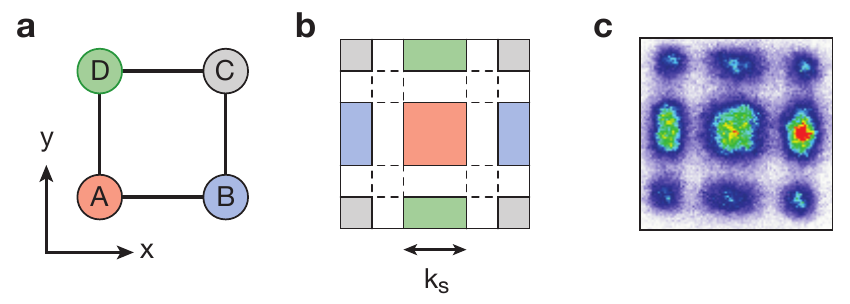}
\caption{
\textbf{(a)} Schematic of the four-site plaquette.
\textbf{(b)} Brillouin zones of the 2D lattice.
\textbf{(c)} Typical momentum distribution obtained after $10\,$ms of time-of-flight.
}\label{fig:S2}
\end{center}
\end{figure}

To detect the occupation numbers on the different sites of the plaquette we apply two mapping sequences along the $x$ and $y$ direction during which the populations $N_q$ are transferred to different Bloch bands analog to the technique described in Ref.~\cite{a.foelling2007direct,a.SebbyStrabley2007atomnumber} for isolated double wells. A subsequent band-mapping technique allows us to determine the population in the Bloch bands by counting the atom numbers in different Brillouin zones. The colors used in Fig.~\ref{fig:S2}(a) and (b) show the connection between the Brillouin zones and the corresponding lattice sites. A typical image obtained after $10\,$ms of time-of-flight is shown in Fig.~\ref{fig:S2}(c).


\begin{thebibliography}{10}
\bibitem{Note}
In trapped systems the degeneracy is slightly lifted and the two lowest energy states are symmetric and anti-symmetric superpositions of the two states.
\end{thebibliography}
\end{document}